\documentclass[]{jfm}

\usepackage{graphicx}
\usepackage{array}
\usepackage{newtxtext}
\usepackage{newtxmath}
\usepackage{natbib}
\usepackage{xcolor}
\usepackage{hyperref}
\usepackage{bm}
\hypersetup{
    colorlinks = true,
    urlcolor   = blue,
    citecolor  = black,
}

\newcommand{\RomanNumeralCaps}[1]
\linenumbers

\title{Three-dimensional velocity gradient statistics in a mesoscale convection laboratory experiment}

\author{Prafulla P. Shevkar,
Roshan J. Samuel,
Christian Cierpka
 \and J\"org Schumacher
   \corresp{\email{joerg.schumacher@tu-ilmenau.de}}}

\affiliation{Institute of Thermodynamics and Fluid Mechanics, Technische Universit\"at Ilmenau, Postfach 100565, D-98684 Ilmenau, Germany}

\begin{document}
\maketitle

\begin{abstract}
We present three-dimensional velocity gradient statistics from Rayleigh–B\'enard convection experiments in a horizontally extended cell of aspect ratio 25, a paradigm for mesoscale convection. The Rayleigh number $Ra$ ranges from $3.7 \times 10^5$ to $4.8 \times 10^6$, the Prandtl number $Pr$ from 5 to 7.1. Spatio-temporally resolved volumetric data are reconstructed from moderately dense Lagrangian particle tracking measurements. All nine components of the velocity gradient tensor from the experiments show good agreement with those from direct numerical simulations, both conducted at $Ra = 1 \times 10^6$ and $Pr = 6.6$. As expected, with increasing $Ra$, the flow in the bulk approaches isotropy in the $x$–$y$ plane. The focus of our analysis is on non-Gaussian velocity gradient statistics. We demonstrate that statistical convergence up to the 6\textsuperscript{th} order is achieved. Specifically, we examine the probability density functions (PDFs) of components of the velocity gradient tensor, vorticity components, kinetic energy dissipation, and local enstrophy at different heights in the bottom half of the cell. The probability of high-amplitude derivatives increases from the bulk to the bottom plate. A similar trend is observed with increasing $Ra$ at fixed height. Both indicate enhanced small-scale intermittency of the velocity field. We also determine derivative skewness and flatness. The PDFs of the derivatives with respect to the horizontal coordinates are found to be more symmetric as the ones with respect to the vertical coordinate. Furthermore, doubly-logarithmic plots of the PDFs of normalised energy dissipation and local enstrophy at all heights show that the left tails follow slopes of 3/2 and 1/2, respectively, in agreement with numerical results. In general, the left tails of the dissipation and local enstrophy distributions show higher probability values with increasing proximity towards the plate, in comparison to those in the bulk. 
\end{abstract}



\section{Introduction}
\label{sec:headings}
Turbulent convection is commonly studied in horizontally extended layers to understand flow dynamics in Earth’s atmosphere, mantle, and oceans, as well as in electronic cooling systems~\citep{Atkinson1996,INCROPERA}. In Rayleigh-Bénard (RB) convection, a fluid is confined between a heated bottom plate and a cooled top plate, resulting in the formation of convection rolls within the bulk of the cell. Turbulence statistics obtained from these convection experiments offer valuable insights into mesoscale geophysical as well as engineering applications \citep{Alam2025}. In fluid turbulence, energy is injected at the largest scales and dissipated at the smallest ones; it is transferred through a broad intermediate range of scales. Therefore, quantifying the rate of kinetic energy dissipation, defined as 
\begin{equation}
    \label{eq:epsi}
\epsilon (x_k,t) = 2\nu {S_{ij}} {S_{ij}}=\frac{\nu}{2} \left( \frac{\partial u_i}{\partial x_j} + \frac{\partial u_j}{\partial x_i} \right)^2,
\end{equation}
is essential for understanding the energy cascade. Here, $\nu$ is the kinematic viscosity of the fluid and $S_{ij}$ the strain rate tensor. Similarly, it is important to characterize the local enstrophy, defined as 
\begin{equation}
\label{eq:omg}
\Omega (x_k,t)= \omega^2 = \frac{1}{2} \left( \frac{\partial u_i}{\partial x_j} - \frac{\partial u_j}{\partial x_i} \right)^2,
\end{equation}
 as it quantifies the intensity of vortical structures across scales, where $\omega$ is the magnitude of the vorticity field $\omega_i=\varepsilon_{ijk} \partial u_k/\partial x_j$, indices $i,j,k=1,2,3$. Both dissipation and local enstrophy exhibit strong and localized fluctuations. The latter has even been found to be more intermittent than the former in homogeneous and isotropic turbulence~\citep{Chen_Srini_Nelkin_1997_inertial_range_scalings,CHEn_Chen_1998,yeung2015_2015_extreme_events}. More broadly, such fluctuations in velocity gradients, vorticity components, enstrophy, and dissipation are collectively referred to as manifestations of intermittency or the vigour of turbulence. The tails of the probability density functions (PDFs) of velocity gradients are known to exhibit a transition from a Gaussian distribution at low Reynolds numbers $Re$ to a non-Gaussian, intermittent nature at higher $Re$ in both homogeneous isotropic turbulence (HIT) and Rayleigh–Bénard convection (RBC)~\citep{Schumacher2014_small_scale_univer}. This transition is evident as increasingly wider tails in the PDFs with increasing Reynolds or Rayleigh number in RBC. Reynolds and Rayleigh number are given by
\begin{equation}
\label{eq:Ra}
\Rey=\frac{U_{\rm rms}H}{\nu}\quad\mbox{and}\quad Ra=\frac{g\alpha\Delta TH^3}{\nu\kappa},
\end{equation}
where $g$ is the acceleration due to gravity, $\Delta T$ the temperature difference between the hot and cold plates, $H$ the height of the convection cell, $\alpha$ the thermal expansion coefficient, $\kappa$ the thermal diffusivity, and $U_{\rm rms}$ the root mean square velocity (see also Sec. 3).  

Furthermore, the transverse component of vorticity, measured using stereoscopic-PIV in RBC experiments with aspect ratio $\Gamma = 10$ in air, also demonstrates transition from Gaussian to non-Gaussian statistics in with respect to velocity derivatives \citep{Valentina_2022_extreme_vorticity}. 
Here, $\Gamma = L/H$, where $L$ is the horizontal length of the cell. \cite{Sharifi2024} showed for the convection cell that out-of-plane vortices are prominent and their number increases with $Ra$ with additional broader tails for the PDFs of the velocity components. The vigour of turbulence in RBC is also characterized by the strength of the internally generated large-scale circulation (LSC), which is known to strongly depend on the aspect ratio of the convection cell~\citep{NIEMELA_SKRBEK_SREENIVASAN_DONNELLY_2001,BAILON-CUBA_EMRAN_SCHUMACHER_2010,wagner_2013,Shevkar_Mohanan_Puthenveettil_2023}. The superstructures, which are evident in high-aspect-ratio cells, are essentially reorganizations of the large-scale flow (LSF) in the bulk. They exhibit stronger fluctuations with increasing $Ra$~\citep{Pandey_superstructures,Moller_Kaufer_Pandey_Schumacher_Cierpka_2022}. This serves as another signature of enhanced intermittency at higher $Ra$.

Numerical approaches enable three-dimensional statistical analyses using both velocity and temperature fields over a broad range of Rayleigh numbers ($Ra$). The Prandtl number $Pr$ is another important control parameter in turbulent convection, where $Pr=\nu/\kappa$. Statistical studies conducted for aspect ratio $\Gamma \sim 1$ and Prandtl number $Pr \sim 1$ are reported in the literature such as in \cite{VERZICCO_CAMUSSI_2003}, \cite{Zhang_Zhou_Sun_2017} and \cite{Vishnu_Mishra2022}. Notably, at the same Grashof number $Gr=Ra/Pr$, a significantly higher flow Reynolds number is observed at $Pr = 0.021$ than at $Pr = 0.7$, indicating more vigorous turbulence at lower Prandtl numbers~\citep{schumacher_2015_EnstrphyGenLowPr}. Performing highly resolved DNS in turbulent RBC for $Pr=0.7$ and $Pr=6$,~\cite{Scheel2013_fine_scale} found that the high-amplitude kinetic energy dissipation and thermal dissipation events and their range of scales decreases with increasing $Pr$. Also, shorter right-tails for the dissipations are observed for the data in the bulk in comparison to that for whole cell. This was also observed by \cite{EMRAN_SCHUMACHER_2008}. This difference in the contribution to high-amplitude events between the bulk and the whole cell further motivates a more detailed statistical study at different heights for various $Ra$ in a high-aspect-ratio channel. 

The high-amplitude tails of the energy dissipation or local enstrophy PDFs, though associated with low-probability events, represent strong fluctuations. Therefore, they have received considerable attention, as they are important for understanding the intermittent and extreme behavior in turbulent flows. These tails asymptotically follow stretched exponential forms~\citep{Donzis_Yeung_Srini2008_stretched_expo,yeung2015_2015_extreme_events,Xu_KQX_2024}. In contrast, the low-amplitude tails, despite being events with a higher probability, are much less studied. Recently, \cite{gotoh2022,gotoh_2023}, using high resolution DNS of isotropic box turbulence, found that the left tail of  kinetic energy dissipation and local enstrophy follow slopes of 3/2 and 1/2, respectively. To the best of our knowledge, this has not yet been reported in controlled laboratory experiments.

Estimating the dissipation or enstrophy using experimental methods such as single-point measurements of Laser Doppler Velocimetry (LDV) or hot-wire anemometry, combined with high temporal sampling rates and Taylor’s hypothesis, are commonly used to estimate the kinetic energy dissipation. Also, two-point, three-point and multi-point LDV~\citep{Ducci_2010_multipointLDV} or hot-wire measurements have been used without relying on Taylor's hypothesis, offering higher accuracy in estimating velocity gradients and dissipation in turbulent flows~\citep{Wallace_2010_hot_wire_3point}. However, since the hot-wire method is intrusive, sensitive to orientation and its spatial resolution depends on the sensor area, it is not always the most suitable approach. Furthermore, non-invasive Particle Image Velocimetry (PIV), two-dimensional or stereoscopic, has been employed to estimate in-plane dissipation or pseudo-dissipation, which is particularly useful in homogeneous isotropic flows~\citep{WANG2021_review}. In addition to computing velocity gradient components, planar PIV has the advantage of detecting flow structures but suffers from issues related to spatial resolution, particularly at high Rayleigh numbers. 

Recently, \cite{Xu_KQX_2024} developed a novel PIV system, termed velocity-gradient-tensor-resolved PIV, designed to capture all nine components of the velocity gradient tensor at a single point. The system employs three lasers and three cameras, each aligned in three mutually orthogonal planes, and was applied to Rayleigh-Bénard convection (RBC) in the range $2 \times 10^8 \le Ra \le 8 \times 10^9$. Using this setup, they found that the pseudo-dissipation near the bottom, center, and sidewall regions with an accuracy of approximately $3\%$, significantly outperforming 1D and 2D surrogate-based approaches. Furthermore, they reported scaling laws for the time-averaged kinetic energy dissipation rate with $Ra$ that show good agreement with the predictions of the Grossmann-Lohse theory~\citep{GROSSMANN_LOHSE_2000,Ahlers2009}. Surprisingly, they found events near the side wall to be highly intermittent compared to those in the bulk and near the bottom plate. The kinetic energy dissipation rate or/and enstrophy has also been extensively studied in various flows, including decaying grid turbulence~\citep{LIBERZON2012_decaying_grid_turb_flow}, dilute polymer solutions~\citep{Liberzon_2006_polymer_solutions} and Taylor-Couette flow~\citep{Tokgoz_etal_JerryW2012_TomoPIV_TaylorCoette}.

In this paper, we present intermittency statistics using all nine components of the velocity gradient tensor, as well as derived quantities such as the vorticity components, enstrophy, and dissipation, in a  $\Gamma=25$ convection cell for $3.7 \times 10^5 \le Ra \le 4.8 \times 10^6$ and $5 \le Pr \le 7.1$. Results are presented at multiple heights in the bottom half of the convection cell. A spatio-temporally resolved 3D velocity field was obtained using the Lagrangian particle tracking method Shake the Box (\textit{STB}) \citep{schanz_2016_shakethebox}. The Lagrangian particle data were further processed using the data reconstruction method \textit{VIC\#} to obtain the Eulerian velocity fields in multiple planes stacked over each other. These planes are parallel to the bottom plate. The spatial resolution of the resulting 3D velocity field is of the order of the Kolmogorov length scale. The PDFs of all nine velocity derivatives are in good agreement with those from DNS at $Ra=1\times10^6$ and $Pr=6.6$. Using the experimental data, we discuss the small-scale intermittency statistics as a function of $Ra$, and compare the bulk statistics with those near the proximity of plate. To this end, we analyse the PDFs of all nine velocity gradients, three components of the vorticity vector, kinetic energy dissipation and local enstrophy. Intermittency increases with increasing $Ra$ and in proximity to the heating plate. Additionally, we compare the left-tail scalings of the kinetic energy dissipation and enstrophy with the recent DNS results by \cite{gotoh2022}. We also discuss other relevant statistical quantities.

The present paper is organised as follows. In \S~\ref{sec:experiments}, we describe the experiments in detail. In \S~\ref{sec:simulations}, we briefly describe the DNS. In \S~\ref{sec:results}, we compare all the nine velocity gradient components from one experiment with that from DNS at same $Ra$ and $Pr$. Further, using experimental results, we discuss the intermittency statistics of velocity gradients and vorticity components at the mid-height (\S~\ref{sec:stat_mid-height}) and in the $z$-direction in the bottom half of the cell (\S~\ref{sec:stat_z_direction}) for various $Ra$. Furthermore, in \S~\ref{sec:dissipation_enstrophy}, we discuss the PDFs of the kinetic energy dissipation and enstrophy and the left-tail scalings for them at various heights. Finally, we summarise the main results in \S~\ref{sec:conclusions}. Appendix~\ref{appA} describes statistical convergence analysis for the velocity gradients at the highest $Ra$.

\section{Experiments}
\label{sec:experiments}
To study the statistics of velocity derivatives, all nine components of the velocity gradient tensor were computed in the Rayleigh-Bénard (RB) convection cell. To this end, we obtain a 3D velocity field using a time-resolved Lagrangian particle tracking velocimetry method \textit{STB} and \textit{VIC\#}, both available in the software package Davis 10.2 (LaVision GmbH). 

\begin{figure}
\centering
  {\includegraphics[width=0.7\linewidth,trim=0 0 0 0,clip]{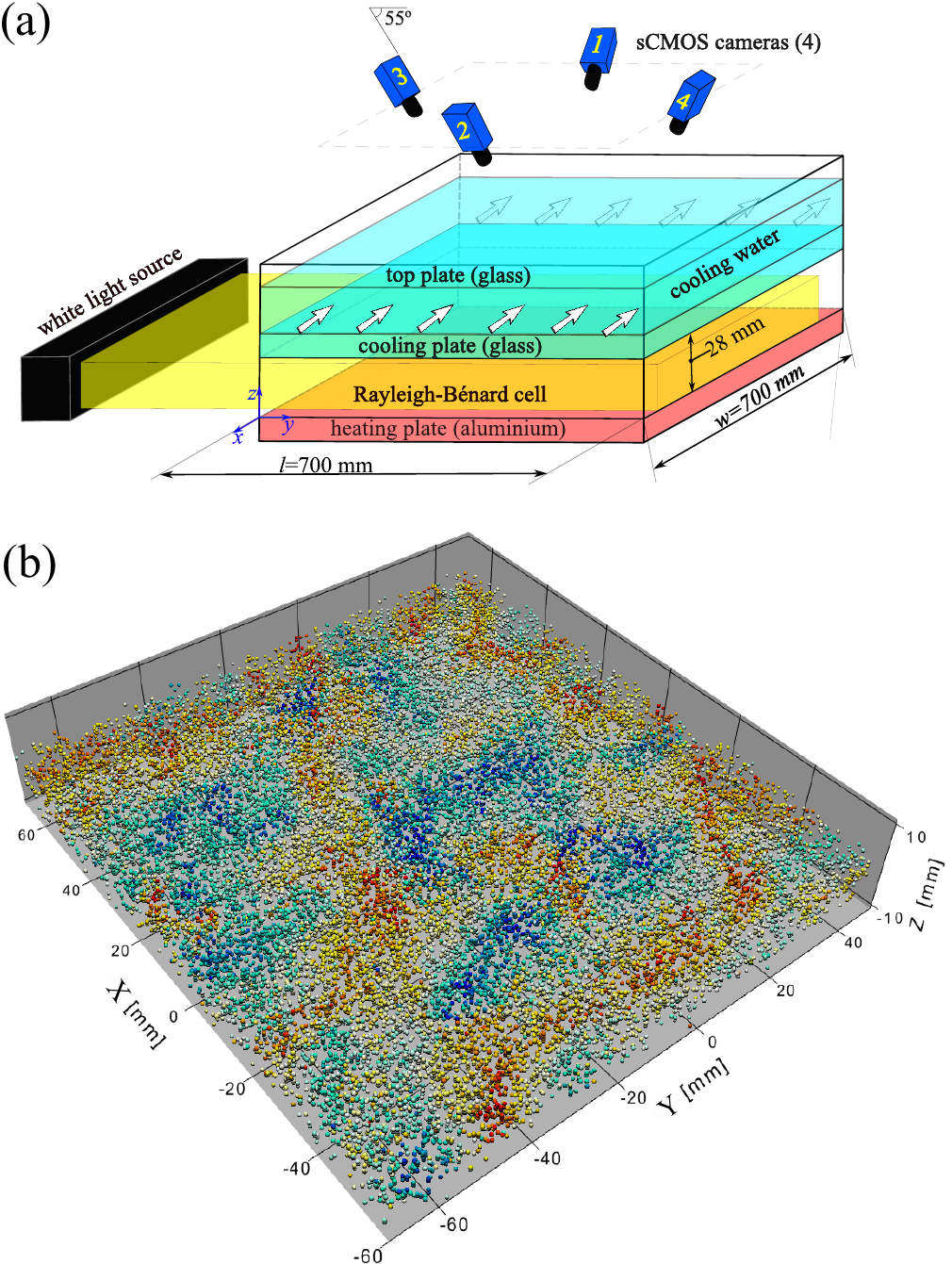}}\\
  \caption{(a) Schematic of the experimental setup of the Rayleigh-B\'enard flow and (b) snapshot showing particles in the measurement volume in the bottom half of the cell at $Ra=4.8\times10^6$. Particles are coloured according to their vertical velocities, with red and yellow indicating positive values, while cyan and blue indicate negative ones. Velocity varies between -6.5 mm s$^{-1}$ and 6 mm s$^{-1}$. The bottom plate is located at $Z=-11$mm. Clearly visible is the turbulent superstructure pattern of the up- and downflows.}
\label{fig:schematic}
\end{figure}

The RBC experiments were conducted in a convection cell with an aspect ratio of $\Gamma = 25$ and a height of $H = 28$ mm. A schematic of the experimental setup is shown in Fig.~\ref{fig:schematic}(a). Water was used as the working fluid, confined between the cooling and heating plates. A constant temperature difference of $\Delta T = T_h - T_c$ was maintained between the heating and cooling plates, where $T_h$ and $T_c$ are the average temperatures of the heating and cooling plates, respectively. The side walls were vertical and adequately insulated to prevent heat loss. Additionally, the room temperature was maintained close to the bulk fluid temperature. The temperature of the cooling plate was measured using four PT-100 thermistors in contact with its surface, while the temperature of the heating plate was measured using four thermistors embedded within it. The average temperature at each plate was maintained constant and used for the calculation of $\Delta T$. The physical properties of water were estimated at the bulk fluid temperature, defined as $T_b = (T_h + T_c)/2$. The heating plate was made of aluminium, and its temperature was kept constant using a constant-temperature bath. The cooling plate was made of glass to allow optical access inside the convection cell. The temperature of the cooling plate was controlled by circulating cold water over it, supplied by a constant-temperature thermostat in a closed-loop system. The temperature difference $\Delta T$ was varied to change the Rayleigh number. The estimated values of $Ra$ and $Pr$ are listed in Table~\ref{tab:expt_parameters}. The maximum deviation in temperature measurements of individual sensors from the mean at the cold and hot plates was estimated. At $Ra = 3.7 \times 10^5$, $9.9 \times 10^5$, and $4.8 \times 10^6$, the deviations at the cold plate were 0.2 K, 0.3 K, and 0.5 K, respectively, while those at the hot plate were 0.06 K, 0.07 K, and 0.5 K, respectively. Except for the highest $Ra$ experiment, the temperature of the hot plate was uniformly maintained within the measurement accuracy. Even though isothermal conditions at the plates are not exactly maintained, the deviations from the mean remained small relative to $\Delta T$. The effects of non-ideal boundary conditions at the plates on the flow dynamics and heat transfer are discussed in~\cite{KAUFER2023283} and \cite{vieweg_2024_digital_twin}. Additional details of the convection cell can be found in~\cite{Moller_Resagk_Cierpka_2020, Moller2021LongtimeEI}.

\begin{table}
  \begin{center}
\def~{\hphantom{0}}
  \begin{tabular}{lccccccccccccc}
      Expt.&$Ra$  & $Pr$   &   $T_h$ &$T_c$& $\Delta T$&fps&FOV&$t_f$&$T_f$&tracked &$\eta_k$&$\Delta x^{Eul}$&PPUC \\[3pt]
        No.&    &        &   \textsuperscript{o}C   & \textsuperscript{o}C & \textsuperscript{o}C    & Hz &mm$\times$mm&s& &particles&mm&mm&particles\\[3pt]
    \textit{Set 1}\\
       1&$3.7\times10^5$   & 7.1 & 20.21 & 18.85&1.36&11&130$\times$125&3.37&540&22 k&1.9&1.1&{0.18}\\
       2&$5.3\times10^5$   & 7.2 & 20.00 & 17.95&2.05&15&130$\times$125&2.82&213&30 k&1.7&1.1&0.25\\
       3&$1.0\times10^6$  & 6.6 & 23.57 & 20.48&3.08&16&130$\times$125&2.11&267&29 k&1.4&1.1&0.24\\
       \textit{Set 2}\\
       4&$3.7\times10^5$   & 7.1 &20.18 & 18.86&1.33&11&130$\times$125&3.40&240&27 k&-&1.4&0.20\\
       5&$9.9\times10^5$   & 6.6 &23.53 & 20.49&3.04&16&130$\times$125&2.12&265&16 k&-&1.7&0.21\\
       6&$4.8\times10^6$   & 5.0 &37.20 & 29.41&7.79&20&130$\times$125&1.08&325&33 k&0.8&1.1&0.13\\
  \end{tabular}
  \caption{Values of the experimental parameters measured. Lagrangian particle tracking measurements were conducted in two sets: Set 1 measurements were taken within a $\pm$5 mm thick illuminated volume around the central $z$-plane, while Set 2 measurements were taken in a 20 mm thick illuminated volume extending from the heating plate into the bulk. The field of view (FOV) is the common field of view of all four cameras. Particles per unit cell (PPUC) refers to the average number of tracked particles within a unit cell, used to determine a velocity vector.}
  \label{tab:expt_parameters}
  \end{center}
\end{table}
The flow was seeded with neutrally buoyant polyamide particles of diameter 55\,\textmu m and density $\rho_p = 1.01$ g cm$^{-3}$. The estimated Stokes number at the highest $Ra = 4.8 \times 10^6$ was $2 \times 10^{-4}$; hence, the particles followed the flow faithfully.
The particles were first mixed thoroughly with water using a magnetic stirrer, and the mixture was allowed to stabilize in a beaker. After approximately half an hour, some particles floated on the water surface, some settled at the bottom, while the neutrally buoyant ones remained suspended. Particle suspension from the central section was then extracted and used for seeding. This method ensured that the floating and settling particles were filtered out, retaining only the neutrally buoyant ones~\citep{Shevkar2019EffectOS}.
 The seeding particles added into the flow were illuminated within a volume using a pulsed white light LED source, see~\cite{Moller_Resagk_Cierpka_2020}. The particle scattering images were captured as a time series using four sCMOS cameras (2560$\times$2160 pixels) from the top of the cell through an imaging window of size 25$\times$25 cm$^2$. The cameras were positioned at an angle of approximately $35^\circ$ to the vertical axis in a symmetric arrangement. All cameras were equipped with 100 mm focal length optics (Zeiss Milvus 2/100 M). Additionally, these lenses were mounted on the cameras via Scheimpflug adapters to optimize the depth of field.

The calibration was carried out in the middle of the cell using a 3D calibration plate (No. 204-15, by LaVision) under actual experimental conditions using a polynomial function. During calibration, the top plane of the plate was positioned at the mid-$z$ plane. The images were masked to a common area imaged by all four cameras. Additionally, the images were pre-processed before performing volume self-calibration~\citep{wieneke2008} and the calibration of particle imaging steps~\citep{Schanz2012_OTF}.
 
Using time-resolved camera images ($1/t_f \ll$ fps) and the generated calibration function, the positions of the particles were computed using \textit{STB}. An instantaneous 3D particle field coloured with their vertical velocities in the bottom half of the cell is shown in Fig.~\ref{fig:schematic}(b). The yellow-red particles, indicating positive vertical velocities, form a connected structure, while the cyan-blue particles, representing negative vertical velocities, form another. Together, most likely they contribute to a larger, organized superstructure in turbulent convection~\citep{Pandey_superstructures,stevens_2018_superstructure,vIEWEG2021,Moller_Kaufer_Pandey_Schumacher_Cierpka_2022,weiss_2023_superstructure,Matti_2025_flow_structures}. The particles were tracked for up to 100 time steps. The total number of tracked particles in the current experimental study for various experiments are listed in Table~\ref{tab:expt_parameters}. 
By fitting a third-order polynomial through the obtained particle positions on a distinct trajectory, the velocities and accelerations of the particles at different locations in the flow field were estimated. A variable time-step processing approach was employed, with an allowed triangulation error of around one voxel. Velocity limits in the $x$- and $y$-directions were set based on the maximum particle displacement per unit time in the horizontal plane, while the limit in the $z$-direction was determined using a trial-and-error method. The acceleration limit was set based on the maximum of the above velocity limits. With manual inspection of the flow field, a decision was made on the use of a spatial filter to remove spurious tracks. Particles not in agreement with the neighbourhood were removed in this step. The allowed triangulation error and the required number of iterations of a spatial median filter during post-processing were optimized to detect true particles while effectively removing spurious tracks. At the highest \( Ra = 4.8 \times 10^6 \), the average uncertainty in the Lagrangian particle velocity magnitude was 13.3\%. This value was calculated from all particles within the measurement volume from a randomly selected 100 consecutive time instants. The distribution of percentage uncertainty was right skewed with a peak at 5.6\%, indicating that most uncertainty values were relatively small. A small fraction of particles displayed high uncertainty values originated from particles with near-zero displacement for any given time step. Similar trends were observed for other two $Ra$ cases.

Furthermore, Eulerian velocity data was obtained by processing particle tracks from \textit{STB} in the \textit{VIC\#} software~\citep{wang_2022_vic_plus,jeon_2022_vicsharp} by solving the cost function for at least 120 iterations. The processing resulted in grid data in the horizontal plane, with several planes stacked on top of each other. Each horizontal plane in Set 1 experiments resulted in a resolution of 1.1 mm with $121 \times 116$ vectors. The vector resolution, $\Delta x^{\text{Eul}}$, on a 3D Eulerian grid for both sets of experiments at various $Ra$ is listed in Table~\ref{tab:expt_parameters}. We also list the Kolmogorov length $\eta_k$ for various $Ra$ using
\begin{equation}
    \label{eq:eta_k}
    \eta_k=\left(\frac{Pr}{Ra}\right)^{3/8}\langle\epsilon\rangle^{-1/4}_{A,t}
\end{equation}
where, $\langle \epsilon\rangle_{A,t}$ is the combined area-time average of the dimensionless kinetic energy dissipation rate at the mid-height~\citep{Scheel2013_fine_scale}. Characteristic units are $\Delta T$ for temperature, $U_f=\sqrt{g\alpha\Delta T H}$ for velocity and $H$ for length, where $U_f$ is the characteristic free fall velocity. The values of the resulting free fall time, $t_f=H/U_f$, and total free fall time, $T_f$, for different experiments are also listed in Table~\ref{tab:expt_parameters}. Based on the number of vectors and total number of tracked particles in a volume for each experiment, particles per unit cell (PPUC) are estimated. In our experiments, PPUC values are ranged in between 0.13-0.25.
\begin{figure}
\centering
  {\includegraphics[width=0.9\linewidth,trim=0 5cm 0 5cm,clip]{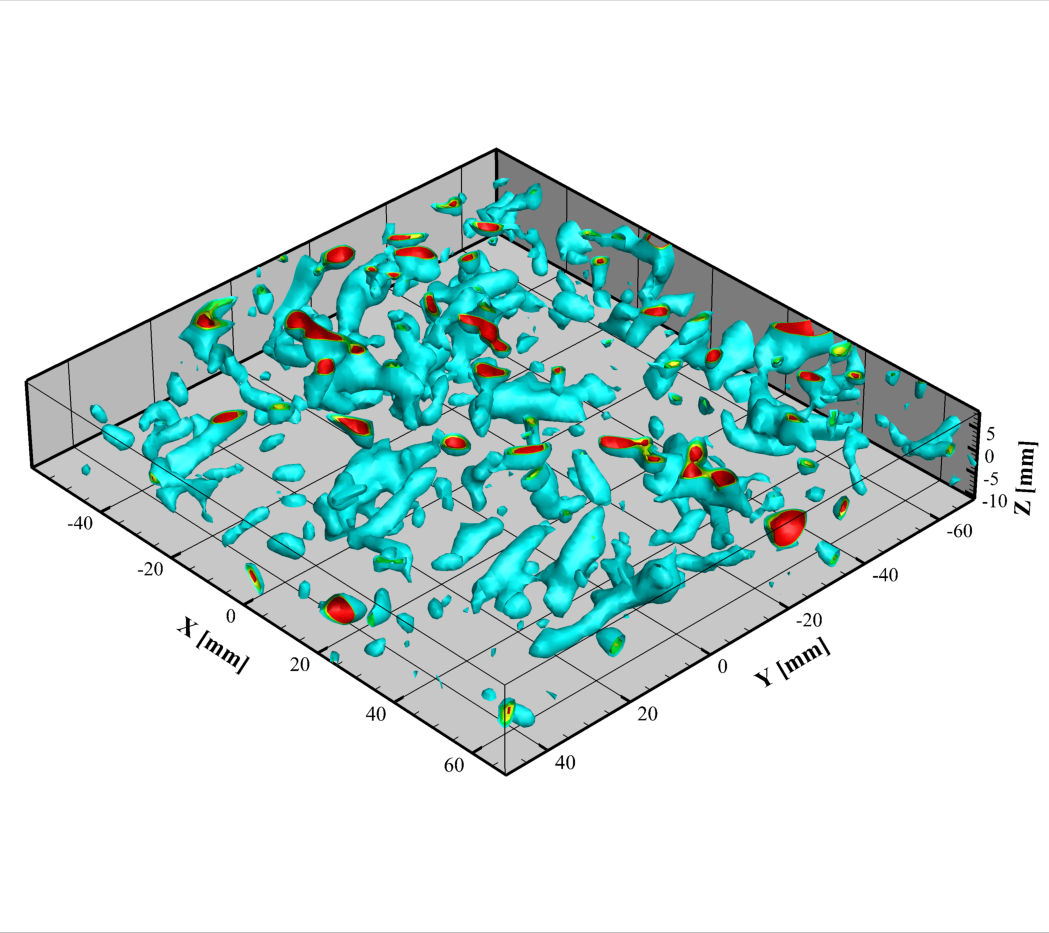}}\\\hspace{1cm}
    \caption{Vortical structures by Q-criterion at a randomly chosen time instance for $Ra=4.8\times10^6$ and $Pr=5$ within the measurement volume. The iso-surfaces of Q within the range of $0.3$ and $0.5$ (cyan-green-yellow-red) are shown.}
\label{fig:structures}
\end{figure}

During each experimental run, temperature control devices, which are connected to the heating and cooling circuits, were set to the desired values. Once a steady state was reached, particles were injected into the flow through a tube connected to the convection cell, and the excess water (about 500 mL) was drained using a tube connected to the other side of the cell. The flow was then allowed to stabilize for half an hour, followed by recordings at the optimal frame rate (fps) for the particle tracking experiments. Figure \ref{fig:structures} shows the isosurfaces of vortical structures at a time instant taken in the RBC experiment. We display isosurfaces in correspondence with the $Q$-criterion \citep{Chong1990}, which are obtained by
\begin{equation}
    \label{eq:Q}
    Q=\frac{1}{2}(O_{ij}O_{ji}-S_{ij}S_{ji})\quad \mbox{with}\quad O_{ij}=\frac{1}{2} \left( \frac{\partial u_i}{\partial x_j} - \frac{\partial u_j}{\partial x_i} 
    \right)\,.
\end{equation}
The tangle of characteristic elongated tube-like structures becomes visible in the bulk, even though Reynolds numbers are relatively low for the present high-Prandtl-number case.   

\section{Direct numerical simulations}
\label{sec:simulations}
We also performed direct numerical simulations for comparison with experiments by solving the three-dimensional Navier-Stokes equations with Boussinesq coupling of velocity ${\bm u}({\bm x},t)=(u_x,u_y,u_z)$ and temperature $T({\bm x},t)$ fields~\citep{Verma2018}. The dimensionless form of the governing equations is given by 
\begin{eqnarray}
\frac{\partial {\bm u}}{\partial t} + ({\bm u} \cdot {\bm \nabla}) {\bm u} & = & -{\bm \nabla}p + T \hat{{\bm z}} + \sqrt{\frac{Pr}{Ra}} \, \nabla^2 {\bm u}, \label{eq:u} \\
\frac{\partial T}{\partial t} + ({\bm u} \cdot {\bm \nabla}) T & = & \frac{1}{\sqrt{PrRa}} \, \nabla^2 T, \label{eq:T} \\
{\bm \nabla} \cdot {\bm u} & = & 0. \label{eq:m}
\end{eqnarray}
The equations are solved by the spectral element method (SEM) using the GPU accelerated SEM solver, nekRS~\citep{Fischer2022}. While the experiments were conducted in a closed cell with aspect ratio $\Gamma = 25$, we choose our computational domain to be a cuboidal cell of $\Gamma = 8$, but with periodic boundary conditions in the horizontal directions. The top and bottom walls enforce no-slip condition on the velocity field and are isothermally maintained at $T_c = 0$ and $T_h = 1$ respectively.

We follow the same workflow as in~\citet{Samuel2024}. Consequently, we ensure that the meshes are fine enough to resolve the boundary layers, and the simulations are run long enough to get reliable statistics for a statistically steady state. The pertinent details of the simulations are listed in Table~\ref{tab:sim_details}. The table also lists the numerically computed Nusselt number, $Nu$, and Reynolds number, $\Rey$, which are the global measures of heat transport and momentum transport respectively. We compute $Nu$ from the non-dimensionalized temperature gradient at the bottom wall, whereas $\Rey$ is calculated from the volume averaged root-mean-square (rms) velocity, see also eq. \eqref{eq:Ra},
\begin{equation}
Nu=-\frac{\partial \langle T\rangle_{A,t}}{\partial z}\bigg|_{z=0},\qquad
\Rey =U_{\rm rms} \sqrt{\frac{Ra}{Pr}}.
\end{equation}
The rms velocity is computed as $U_{\rm rms}=\langle {\bm u}^2\rangle_{V,t}^{1/2}$. The standard deviations of $Nu$ and $\Rey$ are computed from their respective time-series and displayed as the measurement uncertainty for each of them in the table.

\begin{table}
  \begin{center}
    \begin{tabular}{ccccccc}
    $Ra$                & $Pr$ & $\Gamma$ &  $N_{\mathrm{pts}}$            & $T_f$ &  $Nu$            & $\Rey$            \\
    $3.7 \times 10^{5}$ & 7.1  & 8        & $1750 \times 1750 \times 448$  & 2750  & $ 6.23 \pm 0.10$ & $ 22.4 \pm  0.2$  \\
    $1.0 \times 10^{6}$ & 6.6  & 8        & $1750 \times 1750 \times 448$  & 2350  & $ 8.49 \pm 0.11$ & $ 41.7 \pm  0.6$  \\
    \end{tabular}
    \caption{
    Details of the direct numerical simulations.
    Listed here are the Rayleigh number, $Ra$, Prandtl number, $Pr$, the aspect ratio, $\Gamma$,
    the number of collocation points along the $x$-, $y$- and $z$-directions, $N_{\mathrm{pts}}$,
    the total averaging time in free-fall units, $T_f$,
    the time-averaged Nusselt numbers at the walls, $Nu$, and the Reynolds number, $\Rey$.
    Mean values in the last two columns are accompanied by their standard deviations.}
    \label{tab:sim_details}
  \end{center}
\end{table}

\section{Results}
\label{sec:results}
In our convection experiments, Eulerian velocity fields were obtained for at least 200 free fall times at all $Ra$, ensuring that the data records are long enough to provide reliable statistics. First, we present the statistical results at the mid-height of the convection cell.
\subsection{Velocity derivative statistics in the midplane}
\label{sec:stat_mid-height}
\begin{figure}
\centering
  {\includegraphics[width=\linewidth,trim=0 0 0 0,clip]{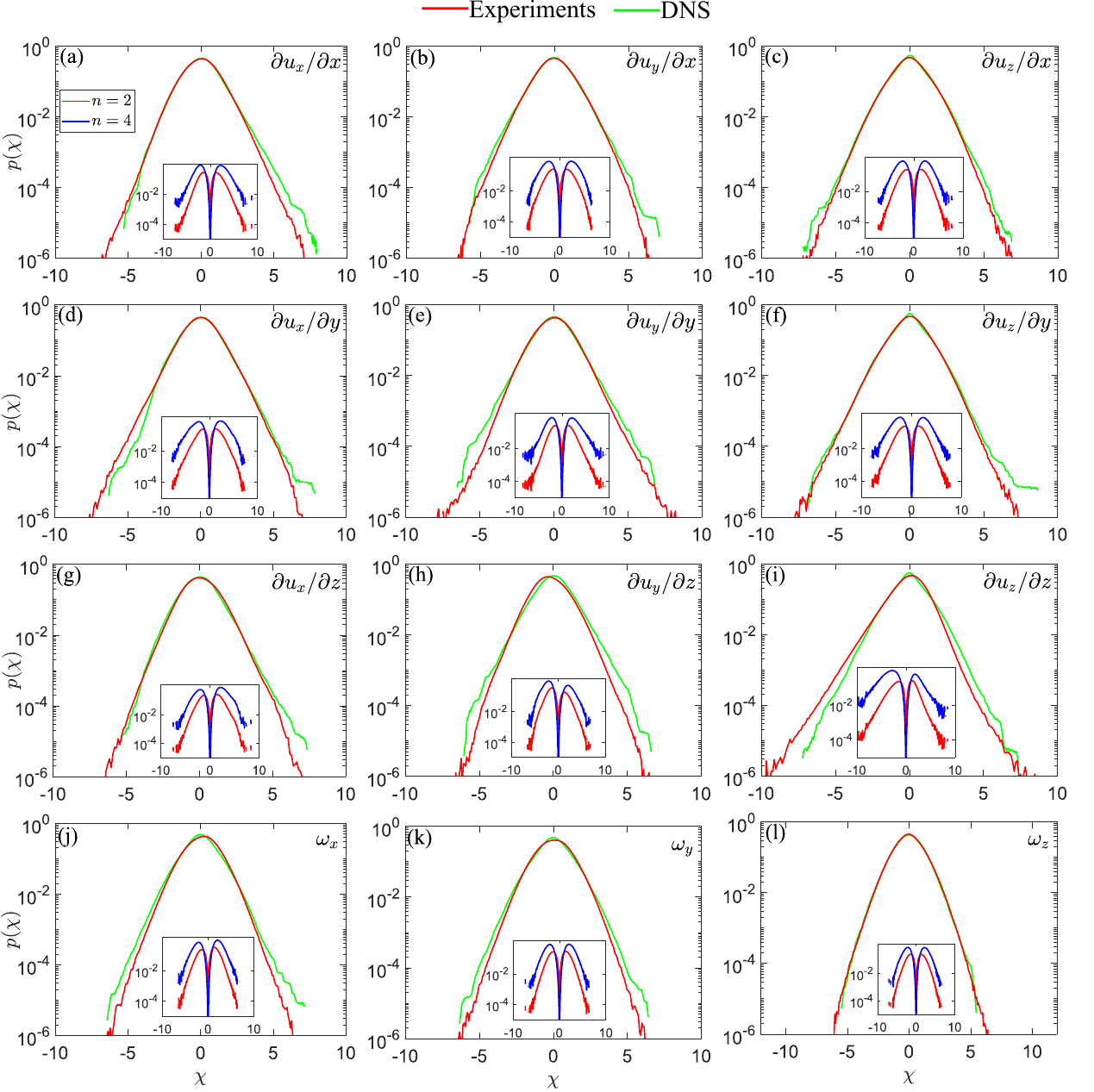}}\\\hspace{1cm}
    \caption{Comparison of probability density functions of the nine velocity gradient tensor components and three vorticity components at the mid-plane for $Ra=1.0\times10^6$ and $Pr=6.6$. The quantities shown are $\partial u_x/\partial x$ in (a), $\partial u_y/\partial x$ in (b), $\partial u_z/\partial x$ in (c), $\partial u_x/\partial y$ in (d), $\partial u_y/\partial y$ in (e), $\partial u_z/\partial y$ in (f), $\partial u_x/\partial z$ in (g), $\partial u_y/\partial z$ in (h), $\partial u_z/\partial z$ in (i),  $\omega_x$ in (j), $\omega_y$ in (k) and  $\omega_z$ in (l). All quantities are normalised by their respective root-mean-square values. Inset figures show the statistical convergences of higher order velocity derivative statistics for the experimental results shown in the main figures. We plot $\chi^n p(\chi)$ versus $\chi$. Here, $n=2$ and 4 for red and blue solid lines, respectively. Note that the y-axis is in logarithmic units.}
\label{fig:ExptVsDNS}
\end{figure}
 The obtained velocity fields are well resolved while also being comparable to the Kolmogorov length scale $\eta_k$. The values of grid resolution and the length scale $\eta_k$ were calculated using \eqref{eq:eta_k} for various $Ra$ at mid-height, are listed in Table~\ref{tab:expt_parameters}. We define 
\begin{equation}
\label{eq:chi}   
\chi = \frac{\omega_i}{\sqrt{\langle \omega_i^2 \rangle_{A,t}}} ~~\text{or}~~ \chi = \frac{\partial u_i / \partial x_j}{\sqrt{\langle \left( \partial u_i / \partial x_j \right)^2 \rangle}_{A,t}} 
\end{equation}
with $i,j = 1, 2, 3$. Note that quantities in \eqref{eq:chi} are normalized by their respective root mean square values. Figure~\ref{fig:ExptVsDNS} compares the experimentally obtained PDFs with those from DNS, obtained in the mid-height for $Ra=1\times10^6$ and $Pr=6.6$. The nine velocity derivatives and three vorticity components are plotted. The experimental data show good agreement with the DNS results, except in some of the far tails.  The good agreement between the experimental data and DNS results reflects the rigor of the experimental measurements and the robustness of the data processing methodology. The insets of Fig.~\ref{fig:ExptVsDNS} demonstrate statistical convergence for the gradient components presented in the respective main figures from the experiments. The decay of higher-order moments towards zero in the log-lin plot of $\chi^n p(\chi)$ vs. $\chi$, for both $n=2$ and $n=4$, indicates sufficient statistical sampling and proper resolution of small-scale structures in the flow. The higher-order moments of $\partial u_z/\partial z$ further unveil highly left skewed nature of $\partial u_z/\partial z$ at the mid-plane, as seen in the inset of Fig.~\ref{fig:ExptVsDNS}(i). This left skewed profile is likely due to the cold and hot plumes plunging into the bulk. Interestingly, gradients of velocities in $z$-direction seem to be more skewed compared to that in $x$- and $y$-directions, which we also discuss using quantitative analysis later in Section~\ref{sec:stat_z_direction}. Similar to $Ra=1\times10^6$, the statistical convergence at the highest $Ra=4.8\times10^6$ is illustrated in appendix~\ref{appA}.

Figure~\ref{fig:intermittent_stats} shows the PDFs of vorticity components in $x$-, $y$- and $z$-directions at the mid-height for four different $Ra$ values, spanning an order of magnitude from the lowest $Ra=3.7\times 10^5$ to the highest $Ra=4.8\times 10^6$ using the experimental data. The Gaussian distributions are also included in the figures as dashed lines for comparison with the PDFs. All the PDFs of vorticity components $\omega_x$, $\omega_y$ and $\omega_z$ exhibit non-Gaussian behaviour with tails which become increasingly wider as $Ra$ number increases. These wider tails in the PDFs of $\omega_i$ are the signatures of increasing small-scale intermittency, similar to non-Gaussian intermittent velocity derivative statistics discussed in~\cite{Schumacher2014_small_scale_univer,Schumacher2018_turb_scal}. Interestingly, the shapes and extents of PDFs of different components of vorticity at same $Ra$ are similar. This suggests that the dominant vorticity structures at mid-height are likely small and do not favor any particular direction for the range of $Ra$, $Pr$ and $\Gamma=25$ convection cells in the current study.   

\begin{figure}
\centering
  {\includegraphics[width=\linewidth,trim=0 0 0 0,clip]{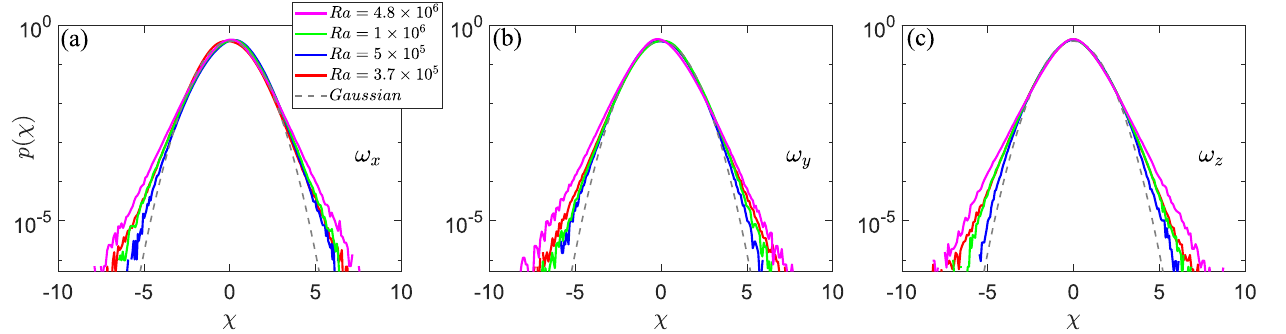}}
    \caption{Probability density functions of vorticity components at the mid-plane for four different $Ra$. The quantities shown are $\omega_x$ in (a), $\omega_y$ in (b) and $\omega_z$ in (c).}
\label{fig:intermittent_stats}
\end{figure}

\subsection{Velocity derivative statistics with respect to $z$-direction}
\label{sec:stat_z_direction}
  \begin{table}
  \begin{center}
\def~{\hphantom{0}}
  \begin{tabular}{lcccccccccccccc}
   \multicolumn{5}{c}{$Ra=3.7\times10^5$} & \multicolumn{5}{c}{$Ra=9.9\times10^5$} & \multicolumn{5}{c}{$Ra=4.8\times10^6$} \\
       $z$   &   $z/\delta_T$ &$G_{11}$&$G_{22}$&$\frac{G_{11}+G_{22}}{2}$  & $z$   &   $z/\delta_T$& $G_{11}$ &$G_{22}$&$\frac{G_{11}+G_{22}}{2}$& $z$   &   $z/\delta_T$ &$G_{11}$&$G_{22}$&$\frac{G_{11}+G_{22}}{2}$ \\[3pt]
          0.51 &7.1&2.00&1.29&1.65 &   0.50 &8.9&1.89&0.94 &1.42&0.48 &13.0&0.87&0.75& 0.81 \\
            0.42 &5.7&2.02&1.35&1.69 &   0.33 &5.8&1.58&0.96&1.27      & 0.21 &5.7&0.94&0.90&0.92  \\
                   0.27 &3.7&1.48&1.33&1.41   & 0.21 &3.7&1.28   &1.09&1.19   & 0.14 &3.6& 1.00&0.96&0.98 \\
            0.17 &2.4&0.74&0.94&0.84   & 0.15 &2.6& 0.88&0.74 &0.81    & 0.10 &2.6&0.90&0.82&0.86  \\        
            0.12 &1.7&0.67&0.69&0.68   & - &-&-& -& -    & - &-&-&- &-\\        
  \end{tabular}
  \caption{List of $x$-$y$ plane locations selected for the analysis and corresponding anisotropy coefficients at three different $Ra$. $\delta_T$ was estimated using an expression of dimensionless heat flux $Nu=0.33Ra^{1/4}Pr^{-1/12}$, see \cite{GROSSMANN_LOHSE_2000}.}
  \label{tab:z_planes}
  \end{center}
\end{table}
\begin{figure}
\centering
  {\includegraphics[width=\linewidth,trim=0 0 0 0,clip]{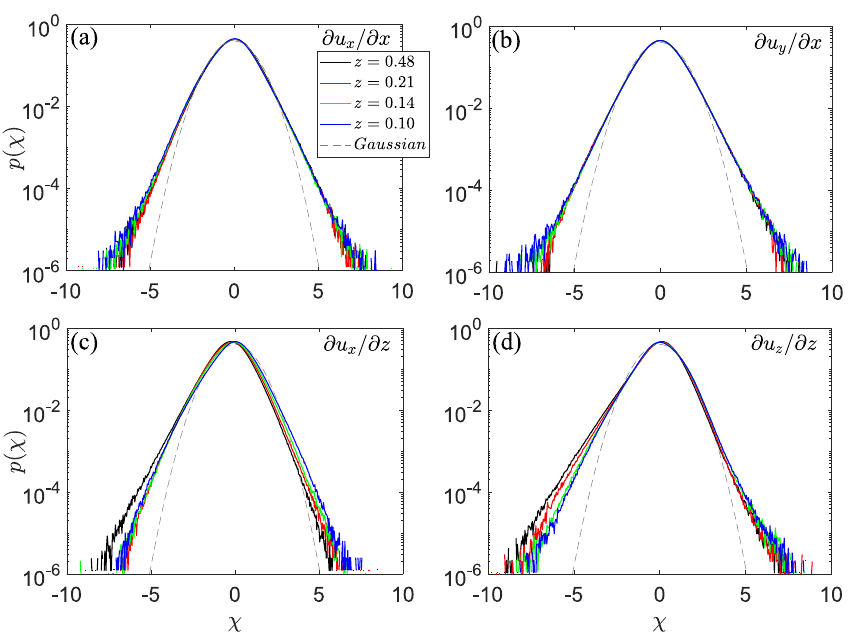}}\\
    \caption{Probability density functions of velocity gradients for four different planes in $z$-direction at the highest $Ra=4.8\times10^6$. The quantities shown are $\partial u_x/\partial x$ in (a), $\partial u_y/\partial x$ in (b), $\partial u_x/\partial z$ in (c) and , $\partial u_z/\partial z$ in (d). $z=0.48$ (black line), $z=0.21$ (red line), $z=0.14$ (green line), $z=0.10$ (blue line), Gaussian reference (dashed, gray line). }
\label{fig:gradient_stats}
\end{figure}
To study derivative statistics in the $z$-direction at each $Ra$, we discuss the PDFs of the velocity derivatives and its derived quantities at four different planes. Details of these planes for three different $Ra$ are given in Table~\ref{tab:z_planes}. These planes are chosen from the bottom half of the convection cell due to the complete availability of data in this region and the symmetry of the flow about mid-height in Rayleigh-B\'enard convection. One of these plane locations is chosen close to the mid-height. The other three planes are chosen such that their height, when normalized by the thermal boundary layer thickness, remains the same for each $Ra$. The values of the normalised heights $z/\delta_T$ of those planes are approximately 2.6, 3.7 and 5.6, are also specified in Table~\ref{tab:z_planes}. This corresponds to ${z}=0.1, 0.14$ and $ 0.21$ at the highest $Ra$.

\subsubsection{Local isotropy in convection flow}
We quantify the degree of anisotropy in second order derivatives using the ratios
\begin{equation}
    \label{e:G_{11}}
    G_{ij}=\frac{\langle (\partial u_i/\partial z)^2\rangle (1+\delta_{iz})}{\langle (\partial u_i/\partial x_j)^2\rangle (1+\delta_{ij})}, \quad\mbox{with}\quad i, j = x, y, z,
\end{equation}
given by \cite{Vorobev2005}. Flows with $G_{ij}=1$ are perfectly isotropic flows while the $G_{ij}\rightarrow0$ are strongly anisotropic. The calculated values of $G_{11}$, $G_{22}$ and the mean of both in our experiments are listed in Table~\ref{tab:z_planes} . These anisotropic coefficients take values close to one at all heights for three $Ra$, indicating slight deviation from isotropy. The deviation in the values among the three considered $Ra$ is highest at the lowest $Ra = 3.7 \times 10^5$, and the flow becomes increasingly isotropic with increasing $Ra$ in the bulk. 

\subsubsection{Velocity gradients and vorticity components}
Figure~\ref{fig:gradient_stats} shows the PDFs of four representative velocity gradient tensor components, two in the $x$-direction and the other two in the $z$-direction, at the highest $Ra = 4.8 \times 10^6$. Firstly, the PDFs at all heights in the bottom half of the convection cell show non-Gaussian intermittent velocity derivative statistics. The PDFs of $\partial u_x/\partial x$ and $\partial u_y/\partial x$ show tails that become somewhat wider as we move closer to the plate, for both the left and right tails of the distribution. This is again a clear sign of increased intermittency near the plate compared to the mid-height. 

In contrast, the right and left tails in the PDFs of $\partial u_x/\partial z$ and $\partial u_z/\partial z$ do not show any systematic behaviour as we move closer to the bottom plate. The right tail becomes increasingly wider as we approach the plate, while the left one shows the opposite behaviour. Partially, this is because the PDFs of $\partial u_z/\partial z$ become less left-skewed as we move toward the plate. The other gradients in the horizontal and vertical directions show similar tail behaviour, as described above in Figs.~\ref{fig:gradient_stats}(a,b) and~\ref{fig:gradient_stats}(c,d), respectively. Similarly, the tail behaviour in the other two lower $Ra$ listed in Table~\ref{tab:z_planes}, follows a similar trend for different planes in the bottom half of the convection cell, as observed at the highest $Ra$.

\begin{figure}
\centering
\includegraphics[width=\linewidth,trim=0 0 0 0,clip]{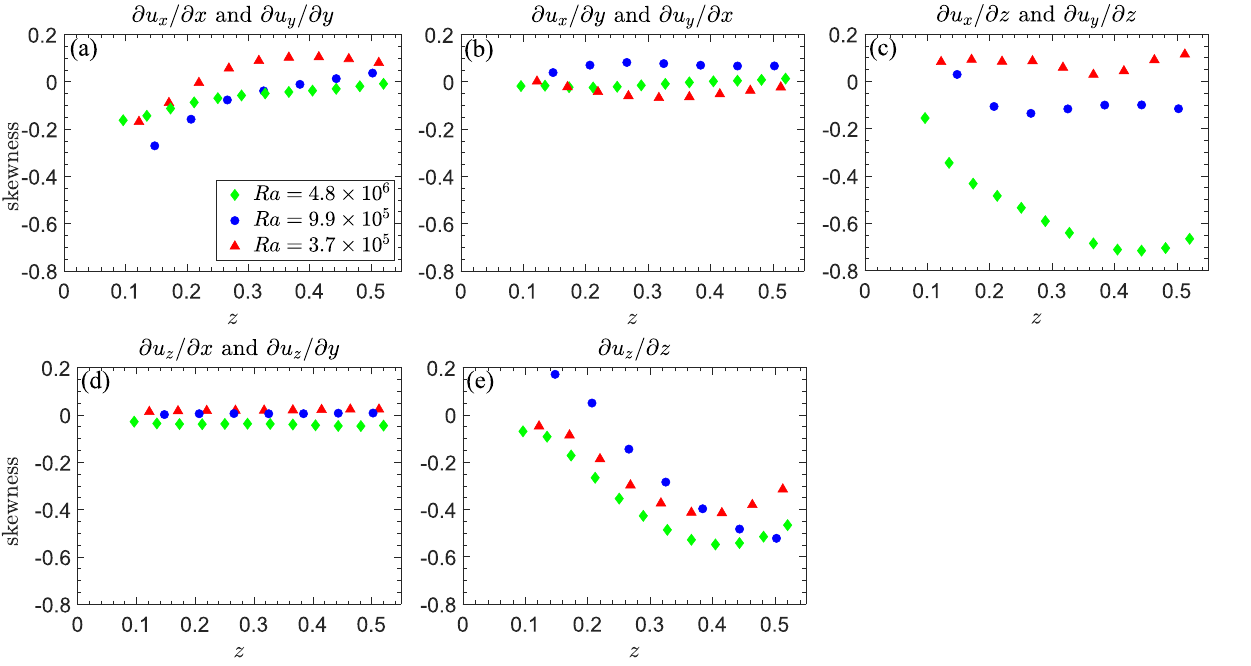}
\caption{Skewness of velocity derivatives, $S_3=\langle\chi^3\rangle/\sigma^3$ at multiple planes in $z$-direction for three different $Ra$. Here, $\sigma$ is standard deviation of $\chi$. The quantities shown are $\partial u_x/\partial x$ and $\partial u_y/\partial y$ in (a), $\partial u_x/\partial y$ and  $\partial u_y/\partial x$ in (b), $\partial u_x/\partial z$ and  $\partial u_y/\partial z$ in (c), $\partial u_z/\partial x$ and  $\partial u_z/\partial y$ in (d) and , $\partial u_z/\partial z$ in (e). Red triangles, $Ra=3.7\times10^5$; blue circles, $Ra=9.9\times10^5$; green diamonds, $Ra=4.8\times10^6$.}
\label{fig:skewness_stats}
\end{figure}
\begin{figure}
\centering
\includegraphics[width=\linewidth,trim=0 0 0 0,clip]{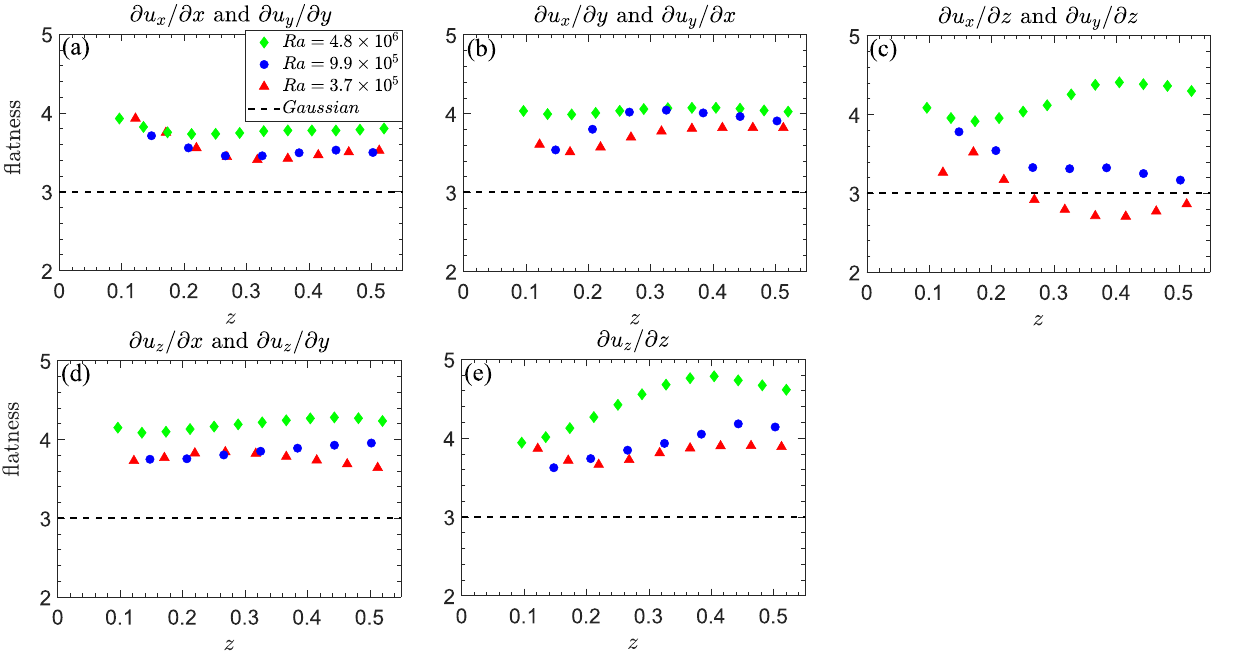}
\caption{Flatness of velocity derivatives, $F_4=\langle\chi^4\rangle/\sigma^4$ at multiple planes in $z$-direction for three different $Ra$. The quantities shown are $\partial u_x/\partial x$ and $\partial u_y/\partial y$ in (a), $\partial u_x/\partial y$ and  $\partial u_y/\partial x$ in (b), $\partial u_x/\partial z$ and  $\partial u_y/\partial z$ in (c), $\partial u_z/\partial x$ and  $\partial u_z/\partial y$ in (d) and , $\partial u_z/\partial z$ in (e). Red triangles, $Ra=3.7\times10^5$; blue circles, $Ra=9.9\times10^5$; green diamonds, $Ra=4.8\times10^6$, $Gaussian$ value of flatness is three (dashed line).}
\label{fig:flatness_stats}
\end{figure}
\begin{table}
  \begin{center}
    \begin{tabular}{ccccccc}
    &$Ra=3.7\times10^5$& $Ra=9.9\times10^5$ &$Ra=4.8\times10^6$ \\
       Skewness $S_3$           &  -0.37$\pm$0.05 &-0.37$\pm$0.16&-0.49$\pm$0.06 \\
       Flatness $F_4$ &3.85$\pm$0.05& 4.03$\pm$0.15& 4.66$\pm$0.13      \\
     
    \end{tabular}
    \caption{The mean values of skewness and flatness for $\partial u_z/\partial z$ in the bulk of the cell, calculated over the height range $0.25 \le z \le 0.5$ along with the corresponding maximum errors.}
    \label{tab:skewz_flatz}
  \end{center}
\end{table}
We calculate the skewness of velocity gradients $S_3=\langle\chi^3\rangle/\sigma^3$ to quantify the asymmetry of their distributions. Figure~\ref{fig:skewness_stats} shows the skewness values of the gradient components in the bottom half of the convection cell at three Rayleigh number values, namely $Ra = 3.7\times10^5$, $9\times10^5$, and $4.8\times10^6$. The velocity derivatives are grouped into five natural pairs  with the same statistics. This grouping is done for symmetry reasons and improves statistical convergence. 

\textit{Skewness of horizontal derivatives.} PDFs of transverse derivatives in $x$- and $y$-directions at different $z$-locations exhibit an approximately symmetric nature, which is shown in Figs.~\ref{fig:skewness_stats}(a,b,d). The skewness values confirm this, as they mostly remain close to zero, ranging between -0.1 and 0.1. The skewness values of the pair of derivatives, $\partial u_z/\partial x$ and $\partial u_z/\partial y$, are independent of $Ra$. In comparison, the skewness values of the longitudinal derivative pair $\partial u_x/\partial x$ and $\partial u_y/\partial y$ at different $Ra$ take increasingly negative values closer to the heating plate, with the highest value approaching -0.3, as seen in Fig.~\ref{fig:skewness_stats}(a). These negatively skewed values for the normal strain components near the heating plate are associated with the events of vigorous plume detachments from the boundary layers close to the heating plate. Plumes are known to contribute to the largest thermal dissipation events near the plate~\citep{Emran_Schumacher_2012}. Furthermore, in the horizontal analysis plane closest to the plate, plumes are the regions of negative horizontal divergence~\citep{Shevkar_plume_detection_2022}. This implies that the flow strongly decelerates in the horizontal direction, likely causing negatively skewed PDFs of normal strains.

\textit{Skewness of vertical derivatives.} The PDFs of derivatives with respect to the $z$-direction are close to symmetric near the heating plate; they become increasingly negatively skewed (and thus asymmetric) deeper into the bulk. Most likely, the descending cold plumes, after detaching from the top wall into the bulk, are responsible for the far left-tail events in those components of the gradient tensor. Hot and cold plumes organize themselves to form large-scale circulations. Thus, this result indicates a reduced influence of the large-scale circulation on the local flow dynamics as we move away from the mid-height towards the plates. The derivative skewnesses in panels (c,e) also show a strong dependence on $Ra$. The pair $\partial u_x / \partial z$ and $\partial u_y / \partial z$ takes skewness values of $S_3= - 0.15$ near the plate which decrease to $-0.7$ at the mid-height for the largest Rayleigh number. Following a similar trend, $\partial u_z / \partial z$ at all three $Ra$ values shows a skewness value approaching $-0.5$ at mid-height, which is typical magnitude of longitudinal derivative skewness of homogeneous isotropic turbulence. Mean values of skewness for the vertical longitudinal derivative $\partial u_z/\partial z$ in the bulk of the cell increases in magnitude from $S_3= - 0.37$ at lowest $Ra=3.7\times10^5$ to $S_3= - 0.49$ at the highest $Ra=4.8\times10^6$, see Table~\ref{tab:skewz_flatz}. These mean values are calculated over the height range of $0.25\le z \le 0.5$. 
\begin{figure}
\centering
  {\includegraphics[width=\linewidth,trim=0 0 0 0,clip]{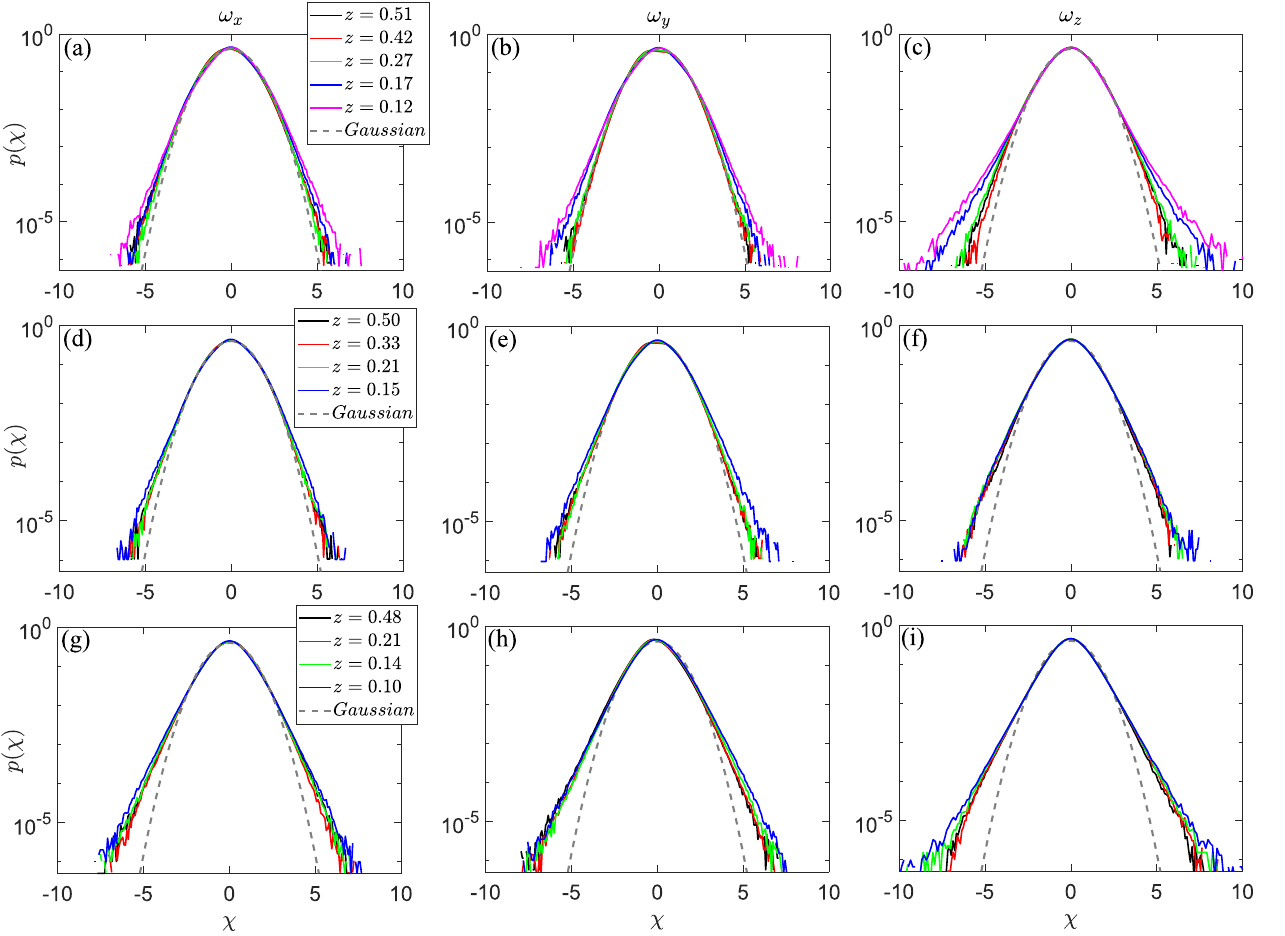}}\\
    \caption{Probability density functions of vorticity components for four different planes in $z$-direction at $Ra=3.7\times10^5$ in (a-c), $Ra=9.9\times10^5$ in (d-f) and $Ra=4.8\times10^6$ in (g-i). The quantities shown are $\omega_x$ in (a, d, g), $\omega_y$ in (b, e, h) and $\omega_z$ in (c, f, i). In (a-c): $z=0.51$ (black), $z=0.42$ (red), $z=0.27$ (green), $z=0.17$ (blue), $z=0.12$ (magenta), $Gaussian$ (dashed, gray). In (d-f): $z=0.50$ (black), $z=0.33$ (red), $z=0.21$ (green), $z=0.15$ (blue), $Gaussian$ (dashed, gray). In (g-i): $z=0.48$ (black), $z=0.21$ (red), $z=0.14$ (green), $z=0.10$ (blue), Gaussian reference (dashed, gray).}
\label{fig:vorticity_stats}
\end{figure}

Similarly, we calculate the flatness values of the velocity gradients by grouping them into five pairs. Figure~\ref{fig:flatness_stats} shows the flatness values at different $z$-planes in the bottom half of the convection cell for three different $Ra$. For a Gaussian field, the flatness takes a value of $F_4=3$. Flatness values greater than three in the bottom half of the convection cell indicate non-Gaussian and intermittent velocity derivative statistics for the range of $Ra$ and $Pr$ considered in the present study.  Increased flatness values with increasing $Ra$ are due to the growing small-scale fluctuations which are associated with higher $Ra$, which was also evident in the tails of vorticity PDFs, as shown previously in Fig.~\ref{fig:intermittent_stats}.

\textit{Flatness of horizontal derivatives}: The variation of the flatness $F_4$ with increasing $Ra$ is small and varies slightly with height for the derivative pairs in Fig.~\ref{fig:flatness_stats}(b,d). However, the statistics for the pair of longitudinal derivatives, $\partial u_x / \partial x$ and $\partial u_y / \partial y$ shows a weak, but systematic increase towards the plate for all three $Ra$, indicating slightly enhanced small-scale intermittency in that region, see panel (a). This is again likely due to plume detachment events.

\textit{Flatness of vertical derivatives}: The increase in flatness values of vertical derivatives with increasing $Ra$ seems to be more significant; in particular the profiles for the largest Rayleigh number can be distinguished from the other data records. A variation at values $F_4 \approx 4$ is observable for $\partial u_x/\partial z$ and $\partial u_y/\partial z$ for $Ra=4.8\times 10^6$ in panel (c). The flatness of $\partial u_z/\partial z$ for this Rayleigh number grows more strongly from the plate into the bulk as seen in panel (e), reaching almost $F_4\approx 5$. For the bulk averages, we find a growth from $F_4=3.85$ at the lowest $Ra=3.7\times10^5$ to 4.66 at the highest $Ra=4.8\times10^6$, see Table~\ref{tab:skewz_flatz}. For all the normalized derivative moments, which we discussed in the last paragraphs, we have to keep in mind that the Rayleigh numbers are moderate; thus the resulting Reynolds numbers for the present $Pr>1$ case lead rather to a three-dimensional time-dependent than a fully developed turbulent flow in the bulk of the convection layer. 

Figure~\ref{fig:vorticity_stats} shows the PDFs of the vorticity components $\omega_x$, $\omega_y$, and $\omega_z$ at multiple $x$-$y$-planes in the bottom half of the convection cell for three different $Ra$ values: $Ra = 3.7 \times 10^5$, $9.9 \times 10^5$, and $4.8 \times 10^6$. Recall that the dimensionless values $z/\delta_T$ for these $x$-$y$-planes are listed in Table~\ref{tab:z_planes}. Previous observations from the PDFs of vorticities at the mid-height in Fig.~\ref{fig:intermittent_stats} demonstrated that intermittency increases with increasing $Ra$. A similar trend is observed for the PDFs of $\omega_x$, $\omega_y$, and $\omega_z$ in Fig.~\ref{fig:vorticity_stats}, where the tails become somewhat fatter with increasing $Ra$. Figure~\ref{fig:vorticity_stats} shows furthermore that the tails of the PDFs, for instance of $\omega_x$, widen as we move closer to the heating plate. This suggests enhanced intermittency near the plate compared to the bulk. The trend holds for all vorticity components, $\omega_x$, $\omega_y$, and $\omega_z$, across the three $Ra$ values. The increased intermittency near the wall is likely caused by turbulent plumes detaching from the boundary layers. This can proceed in the form of swirling motion. 
 
\subsection{Statistics of energy dissipation and local enstrophy}
\label{sec:dissipation_enstrophy}
\begin{figure}
\centering
  {\includegraphics[width=\linewidth,trim=0 0 0 0,clip]{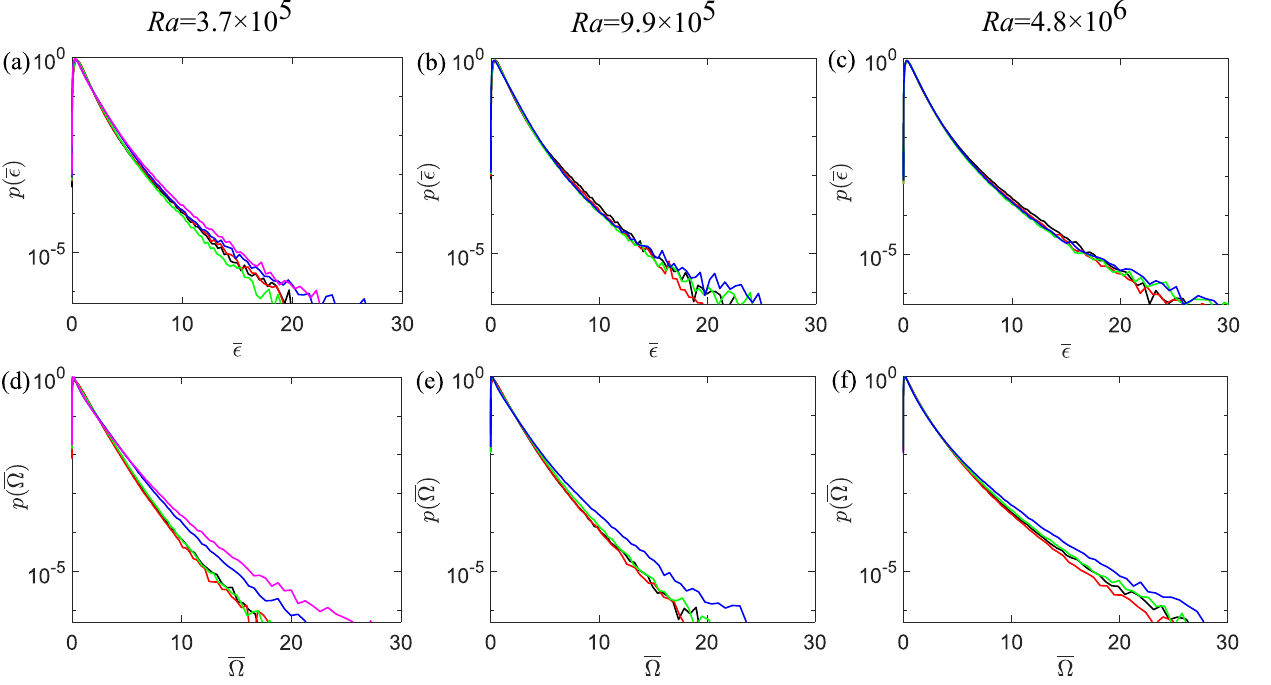}}\\\hspace{1cm}
    \caption{Probability density functions of normalised kinetic energy dissipation rate (top row) and local enstrophy (bottom row) at multiple planes in $z$-direction at $Ra=3.7\times10^5$ in (a,~d), $Ra=9.9\times10^5$ in (b,~e) and $Ra=4.8\times10^6$ in (c,~f). $\overline{\epsilon}=\epsilon/\langle\epsilon\rangle_{A,t}$; $\overline{\Omega}=\Omega/\langle\Omega\rangle_{A,t}$. The heights of planes in $z$-direction are indicated in same way as that in Fig.~\ref{fig:vorticity_stats}. In (a,d): $z=0.51$ (black), $z=0.42$ (red), $z=0.27$ (green), $z=0.17$ (blue), $z=0.12$ (magenta), Gaussian reference (dashed, gray). In (b,e): $z=0.50$ (black), $z=0.33$ (red), $z=0.21$ (green), $z=0.15$ (blue), Gaussian (dashed, gray). In (c,f): $z=0.48$ (black), $z=0.21$ (red), $z=0.14$ (green), $z=0.10$ (blue), Gaussian  reference (dashed, gray).}
\label{fig:eps_ens_stats}
\end{figure}
\begin{figure}
\centering
  {\includegraphics[width=\linewidth,trim=0 0 0 0,clip]{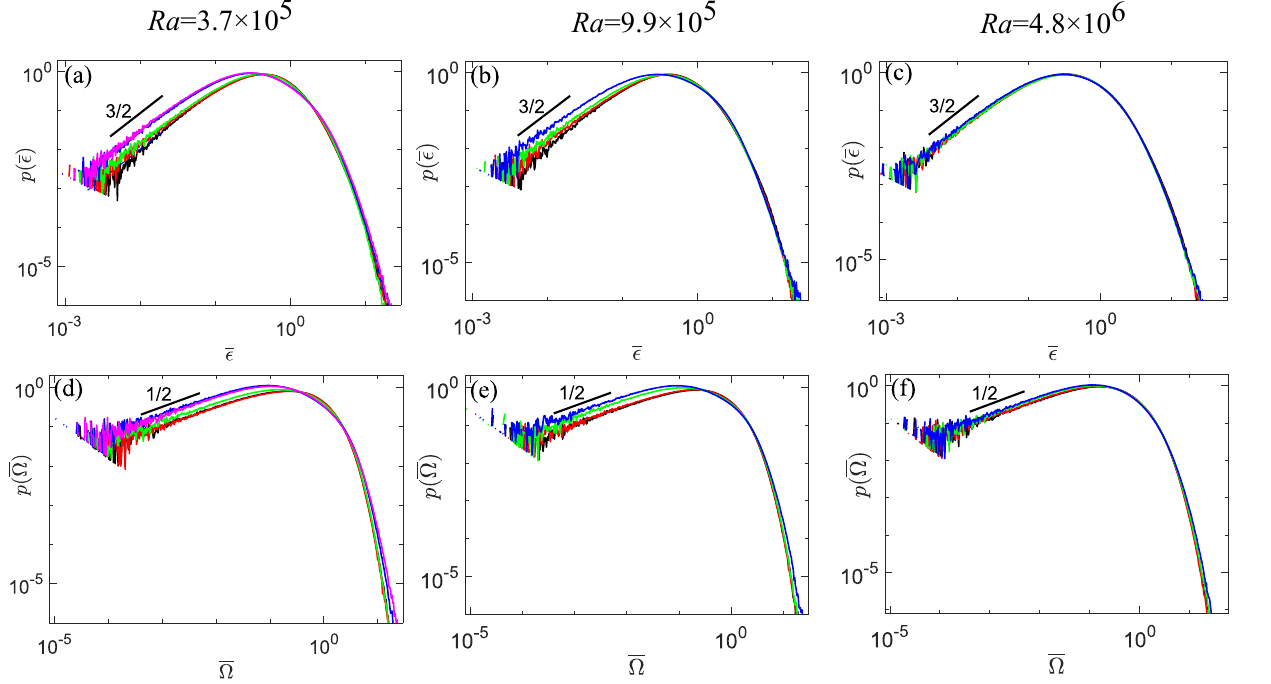}}\\\hspace{1cm}
    \caption{Figures are same as those in Fig.~\ref{fig:eps_ens_stats} except both $x$ and $y$ axes are in logarithmic units. In (a,d): $z=0.51$ (black), $z=0.42$ (red), $z=0.27$ (green), $z=0.17$ (blue), $z=0.12$ (magenta), $Gaussian$ (dashed, gray). In (b,e): $z=0.50$ (black), $z=0.33$ (red), $z=0.21$ (green), $z=0.15$ (blue), Gaussian reference (dashed, gray). In (c,f): $z=0.48$ (black), $z=0.21$ (red), $z=0.14$ (green), $z=0.10$ (blue), Gaussian (dashed, gray).}
\label{fig:eps_ens_loglog_stats}
\end{figure}
\begin{table}
  \begin{center}
\def~{\hphantom{0}}
  \begin{tabular}{lcccccccc}
    \multicolumn{3}{c}{$Ra=3.7\times10^5$} & \multicolumn{3}{c}{$Ra=9.9\times10^5$} & \multicolumn{3}{c}{$Ra=4.8\times10^6$} \\ 
      $z$   &   $\langle \epsilon\rangle_{A,t}$&$\langle \Omega\rangle_{A,t}$& $z$   &   $\langle \epsilon\rangle_{A,t}$&$\langle \Omega\rangle_{A,t}$ & $z$   &  $\langle \epsilon\rangle_{A,t}$&$\langle \Omega\rangle_{A,t}$  \\
       0.51  & $3.8\times10^{-3}$ &0.95&0.50  & $2.2\times10^{-3}$ &0.98&    0.52  & $1.8\times10^{-3}$ & 1.86 \\
         0.46  & $3.6\times10^{-3}$ & 0.92 &    0.44 & $2.2\times10^{-3}$ & 0.96 &    0.48  & $1.8\times10^{-3}$ & 1.80 \\       
         0.42  & $3.4\times10^{-3}$ & 0.90 &   0.38  & $2.1\times10^{-3}$ & 0.94 &     0.44 & $1.7\times10^{-3}$ & 1.76 \\       
         0.37  & $3.3\times10^{-3}$ & 0.87 &   0.33  & $2.1\times10^{-3}$ & 0.93 &     0.41 & $1.7\times10^{-3}$ & 1.75 \\       
         0.32  & $3.2\times10^{-3}$ & 0.83 &   0.27  & $2.0\times10^{-3}$ & 0.94 &    0.37  & $1.7\times10^{-3}$ & 1.75 \\       
         0.27  & $3.1\times10^{-3}$ & 0.81 &   0.21  & $1.8\times10^{-3}$ & 0.89 &    0.33  & $1.7\times10^{-3}$ & 1.77 \\       
         0.22  & $3.0\times10^{-3}$ & 0.79 &   0.15  & $1.8\times10^{-3}$ & 0.72 &   0.29  & $1.7\times10^{-3}$ & 1.79 \\       
         0.17  & $3.0\times10^{-3}$ & 0.73 &     &  & &    0.25  & $1.7\times10^{-3}$ & 1.82 \\       
         0.12  & $3.4\times10^{-3}$ & 0.63 &     &  &  &    0.21  & $1.7\times10^{-3}$ & 1.86 \\       
           &  &     &  &  &   &  0.17 & $1.8\times10^{-3}$ & 1.94 \\       
           &  &     &  &  &   & 0.14  & $1.8\times10^{-3}$ & 1.97 \\       
           &  &     &  &  &   & 0.10  & $1.9\times10^{-3}$ & 1.86
        \\        
  \end{tabular}
   \caption{Mean kinetic energy dissipation rate and enstrophy at multiple $z$-planes in the bottom half of the convection cell for three different $Ra$.}
  \label{tab:planes_dissipation_enstrophy}
  \end{center}
\end{table} 
Figure~\ref{fig:eps_ens_stats} shows PDFs of the normalised kinetic energy dissipation rate and local enstrophy at multiple $x$-$y$-planes in the bottom half of the convection cell, again for three different values $Ra = 3.7 \times 10^5$, $9.9 \times 10^5$, and $4.8 \times 10^6$. The heights of these planes are the same as those in Figure~\ref{fig:vorticity_stats}. A comparison of the right tails of the dissipation and local enstrophy PDFs confirms that intermittency intensifies with increasing $Ra$. Next, we examine the right tails of the dissipation and enstrophy PDFs at mid-height and three additional planes located at the same heights $z/\delta_T$, such that they are comparable. These data are represented by exactly the same line style as that shown in legends of Fig.~\ref{fig:vorticity_stats}. The tails extend to higher values of dissipation and local enstrophy as the planes approach the bottom plate, indicating increased intermittency near the plate. Additionally, for $Ra = 3.7 \times 10^5$, we include magenta curves in the plots, representing the plane closest to the bottom plate. This further reinforces the idea that intermittency increases as we move closer to the bottom plate. This behaviour follows the trend observed in the vorticity PDFs, which also show increased intermittency as we approach the bottom plate. 

In Fig. \ref{fig:eps_ens_loglog_stats}, we replot the data of Fig. \ref{fig:eps_ens_stats}, but on a doubly logarithmic scale, to examine the left tail behaviour. Interestingly, we observe a clear power law scaling for the left tail, with a 3/2 slope for the dissipation and a 1/2 slope for the enstrophy. This provides experimental confirmation of the left-tail scaling trends in convection which were previously suggested by \cite{gotoh2022} for homogeneous isotropic turbulence. The scaling thus remains valid for a large range of Reynolds numbers in different turbulent flows. The probability of left-tail events of the dissipation or enstophy is generally higher for the near the wall regions compared to the bulk. As $Ra$ increases, the tails of energy dissipation and local enstrophy at different heights approach each other, but convergence is not yet evident within the current $Ra$ range. Further, we list the kinetic energy dissipation and enstrophy values for various heights at three different $Ra$ in Table~\ref{tab:planes_dissipation_enstrophy}. At any specific height, the spatio-temporal averaged values of dissipation decrease with an increase in $Ra$, while those of enstrophy increase. The values of dissipation and local enstrophy are found to remain almost constant in the bulk of the convection cell.

\section{Conclusions and outlook}
\label{sec:conclusions}
We presented results on the statistics of components of the velocity gradient tensor from spatio-temporally resolved 3D velocity fields in Rayleigh-Bénard convection experiments. These experiments were conducted in a cell with square cross section with an aspect ratio of 25, a simple laboratory setup of mesoscale convection. The results were presented at mid-height and multiple planes in the bottom half of the convection cell. They covered a parameter range $3.7 \times 10^5 \le Ra \le 4.8 \times 10^6$ and $5 \le Pr \le 7.1$.  

Moderately dense particle images ($\sim$0.01 particles per pixel) were recorded using four sCMOS cameras. The particle images were further processed using the Lagrangian particle tracking velocimetry algorithm known as \textit{Shake-The-Box}, available in the software package DaVis 10.2. The resulting particle positions, velocities, and accelerations were then processed to obtain complete three-dimensional Eulerian velocity fields using \textit{VIC\#}. The final spatial resolution of velocity fields is of the order of the Kolmogorov length scale. At mid-height, all nine components of the velocity gradient tensor, as well as the components of vorticity obtained from one series of the experiments, were in good agreement with the ones from DNS at the same $Ra = 1 \times 10^6$ and $Pr = 6.6$. Along with this, the statistical convergence of the derivatives up to 6th order supports the reliability of the far-tail gradient events and the accuracy of the experimental data reported in the present study. 

Our study was focused on derivative statistics. To this end, we examined the high-amplitude events or tails of the PDFs of all nine components of the velocity gradient tensor, three vorticity components, kinetic energy dissipation, and local enstrophy in four planes located in the bottom half of the convection cell. The planes at three different $Ra$, other than the one at the mid-height, were chosen such that their normalized vertical location, $z/\delta_T$, was the same. This choice enabled a direct comparison of planes located at the same normalized height for different $Ra$ values. The tails of the PDFs of the derivatives become increasingly wider with increasing $Ra$, consistent with \cite{Schumacher2014_small_scale_univer}. This was further confirmed by the systematic increase in flatness values with increasing $Ra$ at all heights in the bottom half of the convection cell. Additionally, the tails also widen as one moves from the bulk region toward the bottom plate, for the range of $Ra$ in the present study. This clearly indicates that intermittency increases both with increasing $Ra$ and in proximity to the plates, compared to the bulk. In general, the estimated skewness values and the shapes of the PDFs of higher-order moments of the velocity gradients indicated that the vertical velocity derivatives are more skewed compared to the horizontal 
ones which might be caused by the thermal plumes that plunge into the central region of the layer. 

Similarly, we also analysed the left tails of the kinetic energy dissipation and local enstrophy at four planes in the bottom half of the convection cell for $Ra=3.7\times10^5,9.9\times10^5 ~\text{and}~4.8\times 10^6$. The left tails display slopes of $3/2$ and $1/2$ at all heights for the kinetic energy dissipation and local enstrophy, respectively. These observed slopes are the experimental confirmation of the highly resolved DNS results by \cite{gotoh2022} for homogeneous isotropic turbulence. Furthermore, in general, these tails exhibited higher probabilities with increasing proximity to the plate, in comparison to the bulk. In the present study, the area-time averaged values of the energy dissipation and local enstrophy remained nearly constant in the bulk of the convection cell. To summarize, our reported laboratory experiments were able to analyse the full three-dimensional small-scale structure of the velocity field in different heights of an RBC layer. 

In the future, we plan to extend the analysis by exploring higher Rayleigh numbers up to $Ra\sim 10^8$. This requires a reduction of the cell aspect ratio to $\Gamma = 10$. The work together with Lagrangian dispersion studies in mesoscale RBC is currently underway and will be reported elsewhere. A further aspect is to combine these volumetric analysis with the temperature field, which can be done for example by a combination with physics-informed machine learning techniques, such as in \cite{Toscano2025}, or using direct measurements~\cite{kaufer_volumetric_2024}.


\backsection[Acknowledgements]{P.P.S. would like to thank Alexander Thieme for the
support with the experiments.}

\backsection[Funding]{The research of P.P.S. and R.J.S. is funded by the European Union (ERC, MesoComp, 101052786). Views and opinions expressed are however those of the authors only and do not necessarily reflect those of the European Union or the European Research Council.}

\backsection[Declaration of interests]{The authors report no conflict of interest.}

\backsection[Data availability statement]{The data that support the findings of this study are available on reasonable request.}

\backsection[Author ORCIDs]{\\
Prafulla P. Shevkar, https://orcid.org/0000-0001-8259-4903 \\
Roshan J. Samuel, https://orcid.org/0000-0002-1280-9881 \\
Christian Cierpka, https://orcid.org/0000-0002-8464-5513\\ 
J{\"o}rg Schumacher, https://orcid.org/0000-0002-1359-4536}

\appendix
\section{Statistical convergence of derivative moments at $Ra=4.8\times10^6$}\label{appA}

\begin{figure}
\centering
  {\includegraphics[width=\linewidth,trim=0 0 0 0,clip]{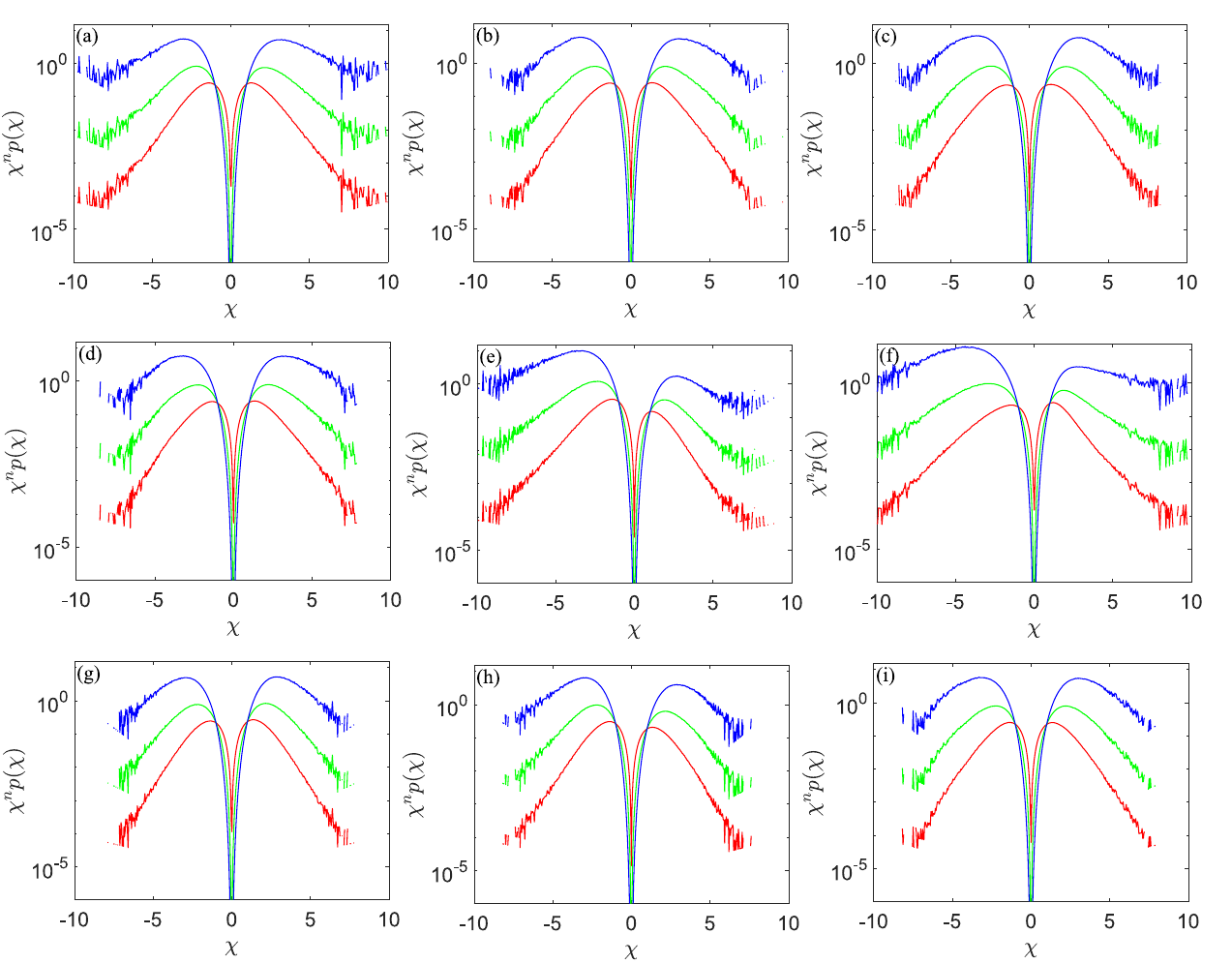}}\\\hspace{1cm}
    \caption{Statistical convergence of higher order velocity derivative-statistics for the experiments at the mid-height for $Ra=4.8\times10^6$ and $Pr=5.0$. The quantities shown are $\partial u_x/\partial x$ in (a), $\partial u_y/\partial x$ in (b), $\partial u_z/\partial x$ in (c), $\partial u_x/\partial y$ in (d), $\partial u_x/\partial z$ in (e), $\partial u_z/\partial z$ in (f),  $\omega_x$ in (g), $\omega_y$ in (h) and  $\omega_z$ in (i). All quantities are normalised by their respective root-mean-square values. Here, $n=2$, 4 and 6 for red, green and blue solid lines, respectively. Note that the y-axis is in logarithmic units.}
\label{fig:stat_convergence2}
\end{figure}

The statistical convergence of the selected velocity derivative and all three vorticity components at the highest $Ra = 4.8 \times 10^6$ at the mid-height is here demonstrated in Fig.~\ref{fig:stat_convergence2}. Here, $n=2$, 4, and 6 stands for the red, green, and blue solid lines in the panels, respectively. The good convergence of the displayed components indicates that the tails tend to decay toward zero. This is definitely obtained for $n=2,4$; a decay is however also observable for $n=6$. Thus we conclude a convergence up to derivative moment order $n=6$. The higher order derivative statistics clearly showcases that $\partial u_x/\partial z$ and $\partial u_z/\partial z$ profiles are more skewed compared to the other velocity derivatives in $x$- and $y$-directions. Similar observations are valid for other three derivative components which are not shown here. This finding also agrees with the numerical and experimental results for RBC in air by \cite{Valori2021} and \cite{Valentina_2022_extreme_vorticity}, respectively.

\bibliographystyle{jfm}
\bibliography{jfm}

\end{document}